\documentclass[lengthcheck,twocolumn,showpacs,10pt,prb,floatfix,amssymb]{revtex4}

\usepackage{graphicx}

\newcommand{\anf}[1]{`#1'}
\newcommand{\abs}[1]{\left|#1\right|}
\newcommand{\vsd}{V_{SD}}

\newcommand{\iv}{$I$-$V$}

\begin{document}
\newlength{\plotwidth}                  
\setlength{\plotwidth}{8.3cm}       
\addtolength{\textheight}{0.5cm}

\title{Shot noise in resonant tunneling through a zero-dimensional state
with a complex energy spectrum}
\author{A.~Nauen}
\email{nauen@nano.uni-hannover.de}
\author{F.~Hohls}
\author{J.~K\"onemann}
\author{R.~J.~Haug}
\affiliation{Institut f\"ur Festk\"orperphysik, Universit\"at Hannover,
Appelstr. 2, D-30167 Hannover, Germany}
\date{\today}
\begin{abstract}
We investigate the noise properties of a GaAs/AlGaAs resonant tunneling structure at bias
voltages where the current characteristic is determined by single electron tunneling. We
discuss the suppression of the shot noise in the framework of a coupled two-state system.
For large bias voltages we observed super-Poissonian shot noise up to values of the Fano
factor $\alpha \approx 10$.
\end{abstract}
 \maketitle

Shot noise allows for a direct measurement of the correlation in a current of discrete
charges. In the case of a totally uncorrelated current, one observed the so called full
or Poissonian shot noise.~\cite{schottky1918} The corresponding noise power $S$ displays
a linear dependence $S=2eI$ on the stationary current $I$ as long as the applied bias
voltage $\vsd$ is large compared to the thermal energy: $e\vsd \gg 2 k_B
T$.~\cite{blanterreview} Full current shot noise arises e.~g.~in a single tunneling
barrier, since the tunneling process of different electrons are independent of each
other.

However, if an additional source of {\it negative} correlation is introduced the noise
amplitude was shown to be reduced.~\cite{li1990,liu1995} For resonant tunneling through a
double barrier structure this is attributed to the dependency of successive tunneling
events caused by the finite dwell time of the resonant state.~\cite{chen1991,davies1992}
This suppression of the shot noise has also been observed for resonant tunneling through
zero-dimensional systems.~\cite{birk1995,nauen2002} For the opposite case of a {\it
positive} correlation between individual tunneling events the noise power can even become
{\it super}-Poissonian.~\cite{iannaccone1998,kuznetsov1998}

We report on noise measurements of a resonant tunneling
double-barrier structure. Under certain bias conditions the
tunneling current through our sample flows through a single
zero-dimensional state.~\cite{schmidt1997,schmidt2001} Therefore
we are dealing with 3d-0d-3d tunneling in contrast to the above
mentioned experimental studies where the tunneling takes place
through a two-dimensional subband
(3d-2d-3d).~\cite{chen1991,davies1992}

\begin{figure}
    \begin{center}
        \includegraphics[width=0.92\linewidth]{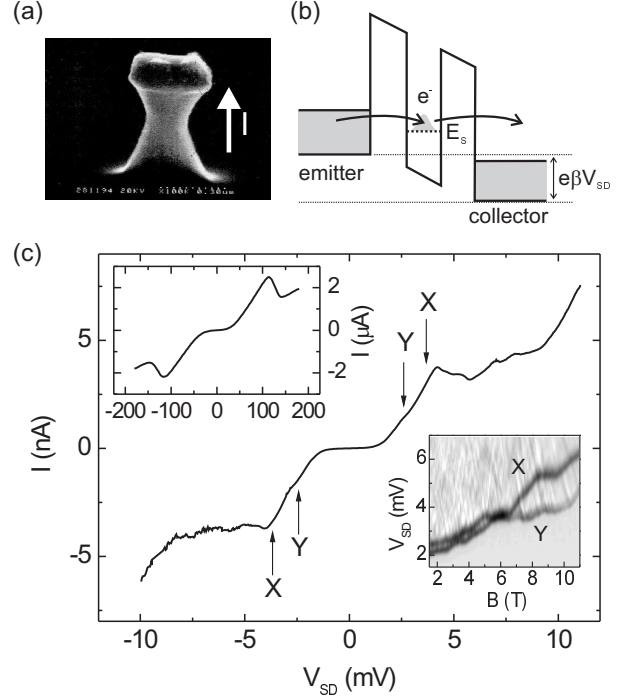}
    \end{center}
    \caption{(a) SEM-Picture of the resonant tunneling diode. (b) Schematical diagram
    of the energy band.
    (c) \iv-characteristic measured at $T=1.3$~K. {\it Upper inset}: \iv-characteristic
    of the sample for larger applied voltages.
    {\it Lower inset}: Differential conductance at a temperature of $T=50$~mK
    in an external magnetic field in parallel to the
    tunneling current.}
    \label{iv-schema}
\end{figure}
Our sample consists of a nearly symmetric double-barrier resonant-tunneling structure
grown by molecular beam epitaxy on a $n^+$-type GaAs substrate (see
Fig.~\ref{iv-schema}a). The heterostructure is formed by a 10 nm wide GaAs quantum well
sandwiched between two Al$_{0.3}$Ga$_{0.7}$As-tunneling barriers of 5 and 6 nm
thickness. The contacts consist of 300~nm thick GaAs layers doped with Si ($4\times
10^{17}~\mbox{cm}^{-3}$) and they are separated from the active region by a 7~nm thick
undoped GaAs spacer layer. The geometrical diameter of the diode is 1~$\mu$m.

The current-voltage (\iv) characteristic is plotted in Fig.~\ref{iv-schema}c.
The upper inset shows the characteristic shape of the resonance due to
tunneling through the two-dimensional subband at $\vsd=\pm 130$~mV. However, we
concentrate our measurements on small bias voltages $\abs{\vsd}\leq 10$~mV,
where we find a pronounced current step (marked with \anf{X}). This feature
indicates the presence of at least one single impurity in the nominally undoped
GaAs, as depicted schematically in Fig.~\ref{iv-schema}b. Such impurity states
with an energy lower than the quantum well in between the barriers most likely
stem from unintended donor atoms within the
well.\cite{dellow92,geim94b,schmidt1997,schmidt2001}

The \iv-curve is slightly asymmetric regarding both bias polarities. We attribute this to
the different transmission coefficients of both tunneling barriers due to their differing
growth thicknesses. In our experiment positive bias voltage means always that tunneling
occurs first through the barrier of lower transmission. Therefore $\vsd
> 0$ corresponds to current flow in non-charging direction since the resonant state is
more often empty during a certain time interval than being occupied. Consequently, $\vsd
< 0$ implies charging direction respectively.

In an external magnetic field in parallel to the direction of the tunneling current $I$
the position of the current step \anf{X} is shifted to higher voltages, overall following
a parabolic dependence as can be seen from the lower inset of Fig.~\ref{iv-schema}c. This
is caused by the diamagnetic shift of the resonant state $E_S$. This allows us to
estimate the classical diameter~\cite{schmidt1997} of the zero-dimensional state to $r
\sim 10$~nm. Additionally, the \iv-curve does reveal a second weak structure which is
marked with \anf{Y} in Fig.~\ref{iv-schema}c. It is more pronounced for charging
direction $(\vsd < 0$). As can be seen in the lower inset of Fig.~\ref{iv-schema}c, in
the magnetospectroscopy this feature first moves parallel to 'X'. At $B=6\,$T both
features merge before splitting again at higher field, but still producing correlated
kinks e.g. at 8 and 10~T. Therefore 'X' and 'Y' are presumably not caused by independent
impurities but stem from a two-level system.~\cite{zhitenev:prl:1997}

The noise experiments are performed in a $^{4}$He bath cryostat with a variable
temperature insert. The sample is always immersed in liquid helium. This allows
for measurements at temperatures between 1.4~K and 4~K under stable conditions.
The bias voltage $V_{SD}$ is applied between the source and drain electrodes by
means of a filtered DC-voltage source. The noise signal is detected by a
low-noise current amplifier.~\cite{nauen:physica2002,nauen2002} The amplifier
output is fed into a voltmeter for measuring the stationary current through the
sample and into a fast Fourier-transform analyzer (FFT) to extract the noise
spectra in a frequency range from 0.4 to 102~kHz with 0.13~kHz resolution. The
amplifier and the input stages of the FFT have been tested for linearity in the
range of interest, overall calibration has been verified by measuring the
thermal noise of thick film resistors.

\begin{figure}
    \begin{center}
        \includegraphics[width=0.7\linewidth]{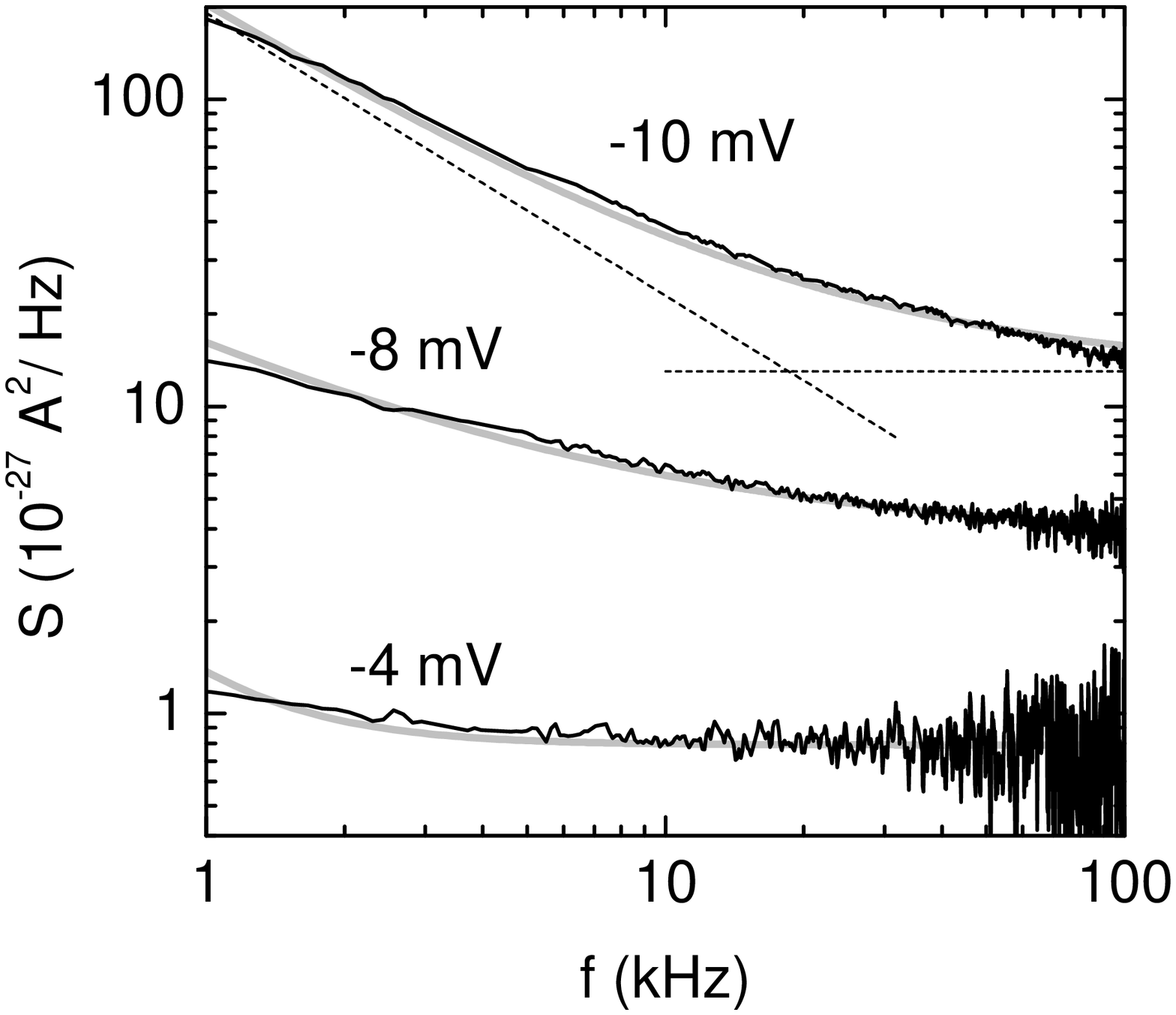}
    \end{center}
    \caption{Measured noise spectra for different values of the bias voltage $\vsd$.
    The grey curves are the results of fitting the function $A/f^\chi+S_0$ to the data.
     The dashed lines demonstrate the contributions of 1/f-noise (angular line) and shot noise
    (horizontal line) to the fit for 10~mV.}
    \label{spektren}
\end{figure}

In Fig.~\ref{spektren} we show exemplarily noise spectra at
different bias voltages for charging-direction. For
$\abs{\vsd}<5$~mV the observed signal is mainly determined by the
frequency independent shot noise. Only below frequencies
$f\lesssim 10$~kHz we find contributions of $1/f$-noise. To
characterize the amplitude of the shot noise we use the so called
Fano factor $\alpha$ which is defined by normalizing the measured
noise power density $S$ to the Poissonian value $2eI$: $\alpha =
S/2eI$ with $e$ the electron charge and $I$ the stationary (DC)
current. For determining $S$ we average the noise spectra over
frequencies larger than the cut-off frequency of the $1/f$-noise.

As the magnitude of the bias voltage is increased, the intensity of $1/f$ noise
rises with an approximate quadratic dependence on current $I$ (data not shown).
At bias voltages above $\abs{\vsd}=8$~mV a direct readout of the Fano factor
becomes impossible since  $1/f$-noise is no longer negligible. However, by
fitting the noise spectra with the function $S(f)=A/f^\chi+S_0$ we can still
extract the amplitude of the shot noise $S_0$: Since the noise processes
leading to shot noise and $1/f$-noise are uncorrelated to each other, their
corresponding noise power add up. Accordingly, the fitting parameter $S_0$ is
the amplitude of the frequency independent shot noise power. For the exponent
$\chi$ characterizing the $1/f$-noise we find $\chi \approx 1$ in agreement
with the literature.~\cite{dutta1981,kirton1989,van-der-ziel}That fit also
reveals that even for the strongest observed 1/f-noise at $V_{sd}=-10$~mV the
shot noise exceeds the 1/f-noise for $f>20$~kHz. Therefore we can extract the
shot noise contribution reliably.

\begin{figure}
    \begin{center}
        \includegraphics[height=0.9\linewidth,angle=270]{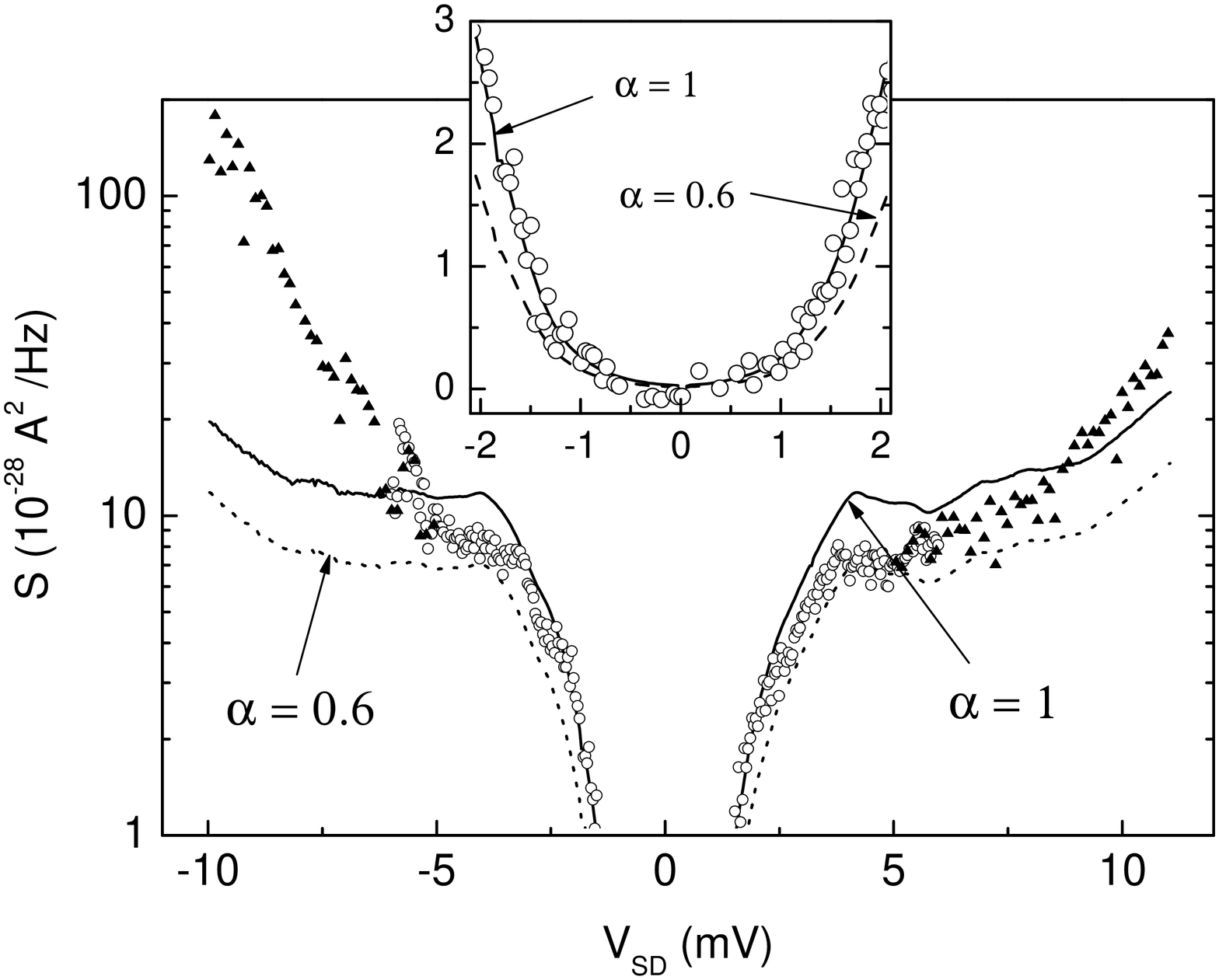}
    \end{center}
    \caption{Shot noise measured as function of bias voltage $V_{SD}$.
    For bias voltages $\abs{\vsd}<6$~mV the shot noise amplitude
    has been defined by averaging the
    spectra over frequencies above the cut-off value of $1/f$-noise (open circles).
    At higher bias voltages a fitting procedure has been used (black triangles, see text).
    {\it Inset}: Shot noise in a linear scale for small bias voltage.}
    \label{rauschen-gesamt}
\end{figure}
In Fig.~\ref{rauschen-gesamt} we plot the measured shot noise
power of the sample for $\abs{\vsd}\leq 10$~mV. Note the good
agreement between the results for the shot noise power extracted
from averaging the spectra (open circles) and from the fitting
procedure (black triangles) in the bias range where both overlap.
Therefore we conclude that the shot noise can be extracted even
under bias conditions where it is masked by $1/f$ noise to a large
extent.

\begin{figure}
    \begin{center}
        \includegraphics[height=\linewidth,angle=270]{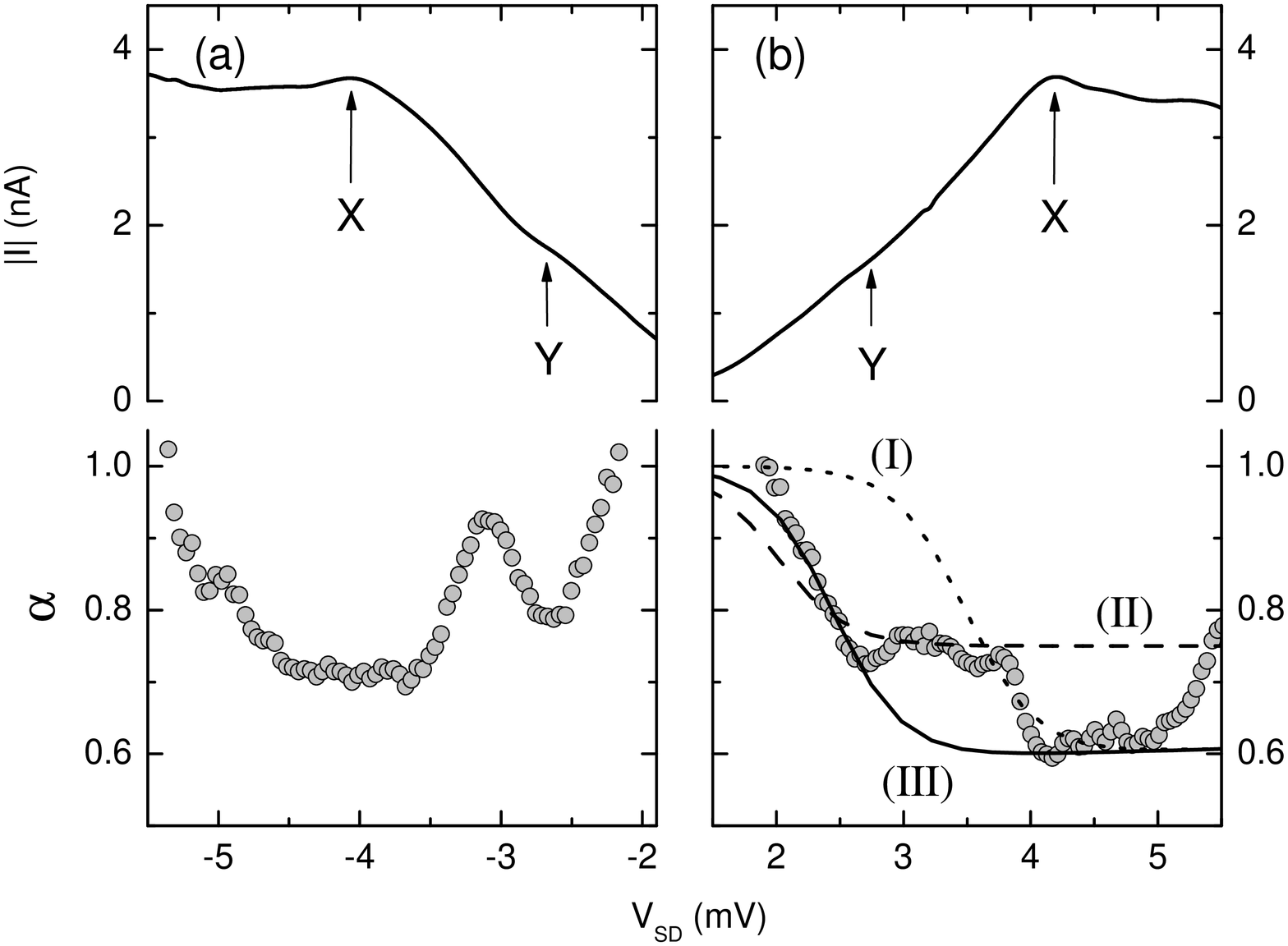}
    \end{center}
    \caption{Behavior of the Fano factor $\alpha$ for the crossover from
        sub-resonant to resonant transport in charging (a) and non-charging direction (b).
        The data has been smoothed with a 7-point
        boxcar average. For the smoothed $\alpha$ we estimate the error to
        $|\delta\alpha| < 0.05$ for $\vsd<3$~mV and
        $|\delta\alpha| < 0.03$ for $\vsd>3$~mV.  For comparison with the data
        we show the theoretically expected characteristics
        under different assumptions (see text).}
    \label{fanosim}
\end{figure}

For comparison we show the theoretically expected Poissonian value for full shot noise
\mbox{($\alpha = 1$)} calculated from the measured DC-current $I$ using \mbox{$S=\alpha
\cdot 2eI\cot\left(e\vsd/2 k_B T\right)$} with the temperature $T$ and Boltmanns's
constant $k_B$.~\cite{blanterreview} As can be seen from the inset in
Fig.~\ref{rauschen-gesamt} we observe full shot noise at bias voltages smaller than
$\abs{\vsd} < 2$~mV. But with increasing absolute value of the bias voltage the noise is
suppressed below its Poissonian value $\alpha = 1$. Overall the suppression is more
pronounced for non-charging direction ($\vsd > 0$). In Fig.~\ref{rauschen-gesamt} also
the theoretically expected noise power for a suppression of $\alpha=0.6$ is shown by the
dashed line. This minimal value of the Fano factor $\alpha \approx 0.6$ is found on the
current plateau $4~\mbox{mV}<\vsd<5\mbox{mV}$. This can be seen in more detail in
Fig.~\ref{fanosim}b, where the Fano factor for non-charging direction is depicted. It has
been shown theoretically that the suppression of shot noise for a resonant state $E_S$ is
linked to the asymmetry of both tunneling barriers~\cite{blanterreview}:
\begin{equation}
        \alpha =
        \frac{\Theta_E^2+\Theta_C^2}{\left(\Theta_E+\Theta_C\right)^2}\;.
        \label{fano}
\end{equation}
$\Theta_E$ and $\Theta_C$ are the tunneling rates through emitter- and collector barrier
respectively. Responsible for this suppression is the Pauli exclusion principle: The
tunneling of an electron from the emitter into the QD is forbidden as long as the QD
state $E_S$ is occupied. This results in an anti-correlation of successive tunneling
events. The maximal suppression of $\alpha = 1/2$ is expected for symmetric barriers
($\Theta_E = \Theta_C$). However, for a tunneling structure of high asymmetry
(e.~g.~$\Theta_E \gg \Theta_C$) full Poissonian shot noise ($\alpha = 1$) would be
recovered, since then the transport will be controlled solely by one barrier.

From the growth data of the sample we can compute the tunneling
rates $\Theta_L$ and $\Theta_C$. This is done by a self-consistent
solution of Poisson- and Schr\"odinger-equation. Using these
values with Eq.~(\ref{fano}) results in $\alpha=0.6$, which is in
good agreement with the experimental result on the pronounced
current step at voltages between 4 and 5~mV (see
Fig.~\ref{fanosim}b).

In Fig.~\ref{fanosim}b the behavior of the Fano factor is plotted for the bias range,
where the crossover into resonant transport occurs. For $\vsd<2$~mV we observe full shot
noise. Above $\vsd=2$~mV the Fano factor is reduced in two steps: For
$3~\mbox{mV}\lesssim \vsd \lesssim 4~\mbox{mV}$ a value of $\alpha\approx 0.75$ is
present. This suppression sets in precisely where a weak resonance can be seen in the
\iv-characteristics (marked with \anf{Y}). For $4~\mbox{mV}\lesssim \vsd \lesssim
5~\mbox{mV}$ the above mentioned value of $\alpha\approx 0.6$ is found, which is
consistent with the growth data of the tunneling structure. Increasing the bias voltage
further results again in a rise of the Fano factor.

It has been shown analytically by Kiesslich et.~al.~\cite{kiesslich2002} that the bias
dependence of $\alpha$ can be described by the expression
\begin{equation}
    \alpha=1-\frac{2\Theta_E\Theta_C}{\left(\Theta_E+\Theta_C\right)^2}f_E\; ,
    \label{fanobias}
\end{equation}
where $f_E^{-1}=1+\exp\left(\left(E_S-\beta e \vsd\right)/k_B T\right)$ is the Fermi
function of the emitter contact (we relate the energies to $E_F=0$) and $\beta$ the
voltage-to-energy conversion factor for the tunneling structure. In case of resonance
and $T=0$ the emitter occupancy at $E=E_S$ is maximal ($f_E(E_S)=1$) and
Eq.~(\ref{fanobias}) reduces to Eq.~(\ref{fano}). At finite temperature $T>0$ the
crossover from Poissonian to sub-Poissonian shot noise is smeared out by the thermal
energy distribution of the emitter electrons.

Now we will discuss the crossover behaviour of the Fano factor as observed in
Fig.~\ref{fanosim}b under different assumptions: At first we consider the features
\anf{X} and \anf{Y} as being caused by tunneling  through two independent resonant
states. For each we expect a crossover to the suppressed shot noise as described by
Eq.~\ref{fanobias}. This is shown by the two curves marked as (I) and (II) in
Fig.~\ref{fanosim}b. In case of (I) we used the calculated values for $\Theta_E(\vsd)$
and $\Theta_C(\vsd)$ and chose a value for $E_S$ so that the crossover from $\alpha=0.75$
to $\alpha=0.6$ coincides with the shape of Eq.~(\ref{fanobias}). For (II) we change the
asymmetry factor to fit a suppression of $\alpha=0.75$. In both cases the crossover of
the measured Fano factor occurs significantly steeper than expected from the form of (I)
and (II) respectively.

However, if we consider just the dominant feature \anf{X} in conjunction with the
calculated tunneling rates for a suppression of $\alpha=0.6$ the crossover from
$\alpha=1$ below $\vsd=2$~mV to $\alpha=0.6$ for $\vsd=4$~mV can be described quite
satisfactorily, of course with a deviation in the bias range $3~\mbox{mV}\lesssim \vsd
\lesssim 4~\mbox{mV}$ (curve (III) in Fig.~\ref{fanosim}b). Thus the behavior of the Fano
factor cannot be explained by two independent states. Instead we conclude, that the
features \anf{X} and \anf{Y} are correlated in a two-fold energy spectrum, in agreement
with our discussion of the magnetotransport spectroscopy shown in Fig.~\ref{iv-schema}c.
The assumption of a correlated two-level system is also stressed by the fact that the
general shape of the Fano factor seems to be mainly determined by a single Fermi-function
((III) in Fig.~\ref{fanosim}b). Whereas the increase of the Fano factor up to $\alpha
\approx 0.75$ around $\vsd \approx 3.5$~mV may be caused by correlation effects on the
current flow due to the two-fold energy spectrum of the resonant state. E.~g.~it has been
shown that a capacitive coupling of two resonant states influences the corresponding shot
noise properties.~\cite{kiesslich2003}

We discuss now the transport in charging direction ($\vsd < 0$): The observed shot noise
is larger compared to non-charging direction as can be seen from Fig.~\ref{fanosim}. In
the theoretical picture for resonant tunneling (Eq.~(\ref{fano}) and (\ref{fanobias}))
the Fano factor $\alpha$ is symmetrical to reversal of bias voltage. However, in a
coupled system of two resonant states of different energies the Fano factor can be
enhanced for charging transport.~\cite{kiesslich2003} We view that as a further
affirmation of the above described perception of a resonant state with a two-fold energy
spectrum.

Finally we will turn to the discussion of the super-Poissonian shot noise
($\alpha > 1$). It is evident from Fig.~\ref{rauschen-gesamt} that for charging
direction $\vsd < 0$ the Fano factor $\alpha$ increases at $\vsd < -6$~mV above
$\alpha=1$ up to values of $\alpha \approx 10$ at $\vsd = -10$~mV. In contrast
this effect is only weakly developed for non-charging direction ($\alpha = 1.5$
at $\vsd = +10$~mV).

Up to now super-Poissonian shot noise has been experimentally
observed for transport through the two-dimensional subband of a
resonant tunneling structure.~\cite{iannaccone1998,kuznetsov1998}
The effect occurs in the negative differential region of the
$I$-$V$-characteristics. It can be explained by a feedback
mechanism that is caused by the charging of the resonant state and
a consequent fluctuation of the subband energy. This leads to a
{\it positive} correlation in the current flow and consequently
the Fano factor becomes super-Poissonian $\alpha >
1$.~\cite{iannaccone1998,blanter1999}

We may speculate that in our sample a similar mechanism occurs although we are dealing
with a zero-dimensional state. This assumption is underlined by the fact that we find
time dependent fluctuations in the $I$-$V$-characteristics for charging direction. These
appear precisely at the bias voltage where the Fano factor crosses over from a sub to a
super-Poissonian value (see Fig.~\ref{iv-schema}c and Fig.~\ref{rauschen-gesamt}). For
non-charging direction we observe no fluctuations of the stationary current and the Fano
factor does not significantly exceed $\alpha=1$.

In conclusion we have analyzed the shot noise properties of
resonant transport through a zero-dimensional state that is formed
by a impurity situated within the quantum well of a tunneling
structure. In the magnetospectroscopy a two-fold energy spectrum
of the resonant state could be identified. The minimal observed
value for the Fano factor of $\alpha=0.6$ does coincide with what
is theoretically expected for the asymmetry of the tunneling
structure as known from the growth data. Although the general
features of the crossover from the full shot noise $\alpha=1$ at
low bias voltage into the suppression of $\alpha=0.6$ can be
satisfactorily described by a single Fermi function, we observe
deviations that we attribute to the complex nature of the resonant
state.

Additionally we observe a super-Poissonian value ($\alpha>1$) of the shot noise for large
bias voltage in charging direction, that is presumably caused by charging effects.

The authors would like to thank G.~Kiesslich for fruitful discussion. We acknowledge
financial support from DFG, BMBF, DIP, and TMR.



\begin{thebibliography}{20}
\expandafter\ifx\csname natexlab\endcsname\relax\def\natexlab#1{#1}\fi
\expandafter\ifx\csname bibnamefont\endcsname\relax
  \def\bibnamefont#1{#1}\fi
\expandafter\ifx\csname bibfnamefont\endcsname\relax
  \def\bibfnamefont#1{#1}\fi
\expandafter\ifx\csname citenamefont\endcsname\relax
  \def\citenamefont#1{#1}\fi
\expandafter\ifx\csname url\endcsname\relax
  \def\url#1{\texttt{#1}}\fi
\expandafter\ifx\csname urlprefix\endcsname\relax\def\urlprefix{URL }\fi
\providecommand{\bibinfo}[2]{#2} \providecommand{\eprint}[2][]{\url{#2}}

\bibitem[{\citenamefont{Schottky}(1918)}]{schottky1918}
\bibinfo{author}{\bibfnamefont{W.}~\bibnamefont{Schottky}},
  \bibinfo{journal}{Ann. d. Phys.} \textbf{\bibinfo{volume}{57}},
  \bibinfo{pages}{541} (\bibinfo{year}{1918}).

\bibitem[{\citenamefont{Blanter and B\"uttiker}(2000)}]{blanterreview}
\bibinfo{author}{\bibfnamefont{Y.~M.} \bibnamefont{Blanter}} \bibnamefont{and}
  \bibinfo{author}{\bibfnamefont{M.}~\bibnamefont{B\"uttiker}},
  \bibinfo{journal}{Phys.~Rep.} \textbf{\bibinfo{volume}{336}}
  (\bibinfo{year}{2000}).

\bibitem[{\citenamefont{Li et~al.}(1988)\citenamefont{Li, Zaslavsky, Tsui,
  Santos, and Shayegan}}]{li1990}
\bibinfo{author}{\bibfnamefont{Yuang~P.} \bibnamefont{Li}},
  \bibinfo{author}{\bibfnamefont{A.}~\bibnamefont{Zaslavsky}},
  \bibinfo{author}{\bibfnamefont{D.~C.} \bibnamefont{Tsui}},
  \bibinfo{author}{\bibfnamefont{M.}~\bibnamefont{Santos}}, \bibnamefont{and}
  \bibinfo{author}{\bibfnamefont{M.}~\bibnamefont{Shayegan}},
  \bibinfo{journal}{Phys. Rev. B} \textbf{\bibinfo{volume}{41}},
  \bibinfo{pages}{8388} (\bibinfo{year}{1988}).

\bibitem[{\citenamefont{Liu et~al.}(1995)\citenamefont{Liu, Li, Aers, Leavens,
  Buchanan, and Wasilewski}}]{liu1995}
\bibinfo{author}{\bibfnamefont{H.~C.} \bibnamefont{Liu}},
  \bibinfo{author}{\bibfnamefont{J.}~\bibnamefont{Li}},
  \bibinfo{author}{\bibfnamefont{G.~C.} \bibnamefont{Aers}},
  \bibinfo{author}{\bibfnamefont{C.~R.} \bibnamefont{Leavens}},
  \bibinfo{author}{\bibfnamefont{M.}~\bibnamefont{Buchanan}}, \bibnamefont{and}
  \bibinfo{author}{\bibfnamefont{Z.~R.} \bibnamefont{Wasilewski}},
  \bibinfo{journal}{Phys. Rev. B} \textbf{\bibinfo{volume}{51}},
  \bibinfo{pages}{5116} (\bibinfo{year}{1995}).

\bibitem[{\citenamefont{Chen and Ting}(1991)}]{chen1991}
\bibinfo{author}{\bibfnamefont{L.~Y.} \bibnamefont{Chen}} \bibnamefont{and}
  \bibinfo{author}{\bibfnamefont{C.~S.} \bibnamefont{Ting}},
  \bibinfo{journal}{Phys. Rev. B} \textbf{\bibinfo{volume}{43}},
  \bibinfo{pages}{R4534} (\bibinfo{year}{1991}).

\bibitem[{\citenamefont{Davies et~al.}(1992)\citenamefont{Davies, Hyldgaard,
  Hershfield, and Wilkins}}]{davies1992}
\bibinfo{author}{\bibfnamefont{J.~H.} \bibnamefont{Davies}},
  \bibinfo{author}{\bibfnamefont{P.}~\bibnamefont{Hyldgaard}},
  \bibinfo{author}{\bibfnamefont{S.}~\bibnamefont{Hershfield}},
  \bibnamefont{and} \bibinfo{author}{\bibfnamefont{J.~W.}
  \bibnamefont{Wilkins}}, \bibinfo{journal}{Phys. Rev. B}
  \textbf{\bibinfo{volume}{46}}, \bibinfo{pages}{9620} (\bibinfo{year}{1992}).

\bibitem[{\citenamefont{Birk et~al.}(1995)\citenamefont{Birk, de~Jong, and
  Sch\"onenberger}}]{birk1995}
\bibinfo{author}{\bibfnamefont{H.}~\bibnamefont{Birk}},
  \bibinfo{author}{\bibfnamefont{M.~J.~M.} \bibnamefont{de~Jong}},
  \bibnamefont{and}
  \bibinfo{author}{\bibfnamefont{C.}~\bibnamefont{Sch\"onenberger}},
  \bibinfo{journal}{Phys. Rev. Lett.} \textbf{\bibinfo{volume}{75}},
  \bibinfo{pages}{1610} (\bibinfo{year}{1995}).

\bibitem[{\citenamefont{Nauen et~al.}(2002{\natexlab{a}})\citenamefont{Nauen,
  Hapke-Wurst, Hohls, Zeitler, Haug, and Pierz}}]{nauen2002}
\bibinfo{author}{\bibfnamefont{A.}~\bibnamefont{Nauen}},
  \bibinfo{author}{\bibfnamefont{I.}~\bibnamefont{Hapke-Wurst}},
  \bibinfo{author}{\bibfnamefont{F.}~\bibnamefont{Hohls}},
  \bibinfo{author}{\bibfnamefont{U.}~\bibnamefont{Zeitler}},
  \bibinfo{author}{\bibfnamefont{R.~J.} \bibnamefont{Haug}}, \bibnamefont{and}
  \bibinfo{author}{\bibfnamefont{K.}~\bibnamefont{Pierz}},
  \bibinfo{journal}{Phys. Rev. B} \textbf{\bibinfo{volume}{66}},
  \bibinfo{pages}{R161303} (\bibinfo{year}{2002}{\natexlab{a}}).

\bibitem[{\citenamefont{Iannaccone et~al.}(1998)\citenamefont{Iannaccone,
  Lombardi, Macucci, and Pellegrini}}]{iannaccone1998}
\bibinfo{author}{\bibfnamefont{G.}~\bibnamefont{Iannaccone}},
  \bibinfo{author}{\bibfnamefont{G.}~\bibnamefont{Lombardi}},
  \bibinfo{author}{\bibfnamefont{M.}~\bibnamefont{Macucci}}, \bibnamefont{and}
  \bibinfo{author}{\bibfnamefont{B.}~\bibnamefont{Pellegrini}},
  \bibinfo{journal}{Phys. Rev. Lett.} \textbf{\bibinfo{volume}{80}},
  \bibinfo{pages}{1054} (\bibinfo{year}{1998}).

\bibitem[{\citenamefont{Kuznetsov et~al.}(1998)\citenamefont{Kuznetsov, Mendez,
  Bruno, and Pham}}]{kuznetsov1998}
\bibinfo{author}{\bibfnamefont{V.~V.} \bibnamefont{Kuznetsov}},
  \bibinfo{author}{\bibfnamefont{E.~E.} \bibnamefont{Mendez}},
  \bibinfo{author}{\bibfnamefont{J.~D.} \bibnamefont{Bruno}}, \bibnamefont{and}
  \bibinfo{author}{\bibfnamefont{J.~T.} \bibnamefont{Pham}},
  \bibinfo{journal}{Phys. Rev. B} \textbf{\bibinfo{volume}{58}},
  \bibinfo{pages}{R10159} (\bibinfo{year}{1998}).

\bibitem[{\citenamefont{Dellow et~al.}(1992)\citenamefont{Dellow, Beton,
  Langerak, Foster, Main, Eaves, Henini, Beaumont, and Wilkinson}}]{dellow92}
\bibinfo{author}{\bibfnamefont{M.~W.} \bibnamefont{Dellow}},
  \bibinfo{author}{\bibfnamefont{P.~H.} \bibnamefont{Beton}},
  \bibinfo{author}{\bibfnamefont{C.~J. G.~M.} \bibnamefont{Langerak}},
  \bibinfo{author}{\bibfnamefont{T.~J.} \bibnamefont{Foster}},
  \bibinfo{author}{\bibfnamefont{P.~C.} \bibnamefont{Main}},
  \bibinfo{author}{\bibfnamefont{L.}~\bibnamefont{Eaves}},
  \bibinfo{author}{\bibfnamefont{M.}~\bibnamefont{Henini}},
  \bibinfo{author}{\bibfnamefont{S.~P.} \bibnamefont{Beaumont}},
  \bibnamefont{and} \bibinfo{author}{\bibfnamefont{C.~D.~W.}
  \bibnamefont{Wilkinson}}, \bibinfo{journal}{Phys. Rev. Lett.}
  \textbf{\bibinfo{volume}{68}}, \bibinfo{pages}{1754} (\bibinfo{year}{1992}).

\bibitem[{\citenamefont{Geim et~al.}(1994)\citenamefont{Geim, Foster, Nogaret,
  Mori, McDonnel, Scala, Main, and Eaves}}]{geim94b}
\bibinfo{author}{\bibfnamefont{A.~K.} \bibnamefont{Geim}},
  \bibinfo{author}{\bibfnamefont{T.~J.} \bibnamefont{Foster}},
  \bibinfo{author}{\bibfnamefont{A.}~\bibnamefont{Nogaret}},
  \bibinfo{author}{\bibfnamefont{N.}~\bibnamefont{Mori}},
  \bibinfo{author}{\bibfnamefont{P.~J.} \bibnamefont{McDonnell}},
  \bibinfo{author}{\bibfnamefont{N.} \bibnamefont{LaScala}},
  \bibinfo{author}{\bibfnamefont{P.~C.} \bibnamefont{Main}}, \bibnamefont{and}
  \bibinfo{author}{\bibfnamefont{L.}~\bibnamefont{Eaves}},
  \bibinfo{journal}{Phys. Rev. B} \textbf{\bibinfo{volume}{50}},
  \bibinfo{pages}{8074} (\bibinfo{year}{1994}).

\bibitem[{\citenamefont{Schmidt et~al.}(1997)\citenamefont{Schmidt, Haug,
  Fal'ko, v.~Klitzing, F\"orster, and L\"uth}}]{schmidt1997}
\bibinfo{author}{\bibfnamefont{T.}~\bibnamefont{Schmidt}},
  \bibinfo{author}{\bibfnamefont{R.~J.} \bibnamefont{Haug}},
  \bibinfo{author}{\bibfnamefont{V.~I.} \bibnamefont{Fal'ko}},
  \bibinfo{author}{\bibfnamefont{K.}~\bibnamefont{v.~Klitzing}},
  \bibinfo{author}{\bibfnamefont{A.}~\bibnamefont{F\"orster}},
  \bibnamefont{and} \bibinfo{author}{\bibfnamefont{H.}~\bibnamefont{L\"uth}},
  \bibinfo{journal}{Phys. Rev. Lett.} \textbf{\bibinfo{volume}{78}},
  \bibinfo{pages}{1540} (\bibinfo{year}{1997}).

\bibitem[{\citenamefont{Schmidt et~al.}(2001)\citenamefont{Schmidt, K\"onig,
  McCann, Fal'ko, and Haug}}]{schmidt2001}
\bibinfo{author}{\bibfnamefont{T.}~\bibnamefont{Schmidt}},
  \bibinfo{author}{\bibfnamefont{P.}~\bibnamefont{K\"onig}},
  \bibinfo{author}{\bibfnamefont{E.}~\bibnamefont{McCann}},
  \bibinfo{author}{\bibfnamefont{V.~I.} \bibnamefont{Fal'ko}},
  \bibnamefont{and} \bibinfo{author}{\bibfnamefont{R.~J.} \bibnamefont{Haug}},
  \bibinfo{journal}{Phys. Rev. Lett.} \textbf{\bibinfo{volume}{86}},
  \bibinfo{pages}{276} (\bibinfo{year}{2001}).

\bibitem[{\citenamefont{Zhitenev et~al.}(1997)\citenamefont{Zhitenev, Ashoori,
  Pfeiffer, and West}}]{zhitenev:prl:1997}
\bibinfo{author}{\bibfnamefont{N.~B.} \bibnamefont{Zhitenev}},
  \bibinfo{author}{\bibfnamefont{R.~C.} \bibnamefont{Ashoori}},
  \bibinfo{author}{\bibfnamefont{L.~N.} \bibnamefont{Pfeiffer}},
  \bibnamefont{and} \bibinfo{author}{\bibfnamefont{K.~W.} \bibnamefont{West}},
  \bibinfo{journal}{Phys. Rev. Lett.} \textbf{\bibinfo{volume}{79}},
  \bibinfo{pages}{2308} (\bibinfo{year}{1997}).

\bibitem[{\citenamefont{Nauen et~al.}(2002{\natexlab{b}})\citenamefont{Nauen,
  K\"onemann, Zeitler, Hohls, and Haug}}]{nauen:physica2002}
\bibinfo{author}{\bibfnamefont{A.}~\bibnamefont{Nauen}},
  \bibinfo{author}{\bibfnamefont{J.}~\bibnamefont{K\"onemann}},
  \bibinfo{author}{\bibfnamefont{U.}~\bibnamefont{Zeitler}},
  \bibinfo{author}{\bibfnamefont{F.}~\bibnamefont{Hohls}}, \bibnamefont{and}
  \bibinfo{author}{\bibfnamefont{R.~J.} \bibnamefont{Haug}},
  \bibinfo{journal}{Physica E} \textbf{\bibinfo{volume}{12}},
  \bibinfo{pages}{865} (\bibinfo{year}{2002}{\natexlab{b}}).

\bibitem[{\citenamefont{Dutta and Horn}(1981)}]{dutta1981}
\bibinfo{author}{\bibfnamefont{P.}~\bibnamefont{Dutta}} \bibnamefont{and}
  \bibinfo{author}{\bibfnamefont{P.~M.} \bibnamefont{Horn}},
  \bibinfo{journal}{Rev. Mod. Phys.} \textbf{\bibinfo{volume}{53}},
  \bibinfo{pages}{497} (\bibinfo{year}{1981}).

\bibitem[{\citenamefont{Kirton and Uren}(1989)}]{kirton1989}
\bibinfo{author}{\bibfnamefont{M.~J.} \bibnamefont{Kirton}} \bibnamefont{and}
  \bibinfo{author}{\bibfnamefont{M.~J.} \bibnamefont{Uren}},
  \bibinfo{journal}{Advances in Physics} \textbf{\bibinfo{volume}{38}},
  \bibinfo{pages}{367} (\bibinfo{year}{1989}).

\bibitem[{\citenamefont{van~der Ziel}(1986)}]{van-der-ziel}
\bibinfo{author}{\bibfnamefont{A.}~\bibnamefont{van~der Ziel}},
  \emph{\bibinfo{title}{Noise in solid state devices and circuits}}
  (\bibinfo{publisher}{John Wiley\&Sons}, \bibinfo{year}{1986}).

\bibitem[{\citenamefont{Kiesslich
  et~al.}(2003{\natexlab{a}})\citenamefont{Kiesslich, Wacker, Sch\"oll, Nauen,
  Hohls, and Haug}}]{kiesslich2002}
\bibinfo{author}{\bibfnamefont{G.}~\bibnamefont{Kiesslich}},
  \bibinfo{author}{\bibfnamefont{A.}~\bibnamefont{Wacker}},
  \bibinfo{author}{\bibfnamefont{E.}~\bibnamefont{Sch\"oll}},
  \bibinfo{author}{\bibfnamefont{A.}~\bibnamefont{Nauen}},
  \bibinfo{author}{\bibfnamefont{F.}~\bibnamefont{Hohls}}, \bibnamefont{and}
  \bibinfo{author}{\bibfnamefont{R.~J.} \bibnamefont{Haug}},
  \bibinfo{journal}{phys.~stat.~sol.~(c)} \textbf{\bibinfo{volume}{0}},
  \bibinfo{pages}{1293} (\bibinfo{year}{2003}{\natexlab{a}}).

\bibitem[{\citenamefont{Kiesslich
  et~al.}(2003{\natexlab{b}})\citenamefont{Kiesslich, Wacker, and
  Sch\"oll}}]{kiesslich2003}
\bibinfo{author}{\bibfnamefont{G.}~\bibnamefont{Kiesslich}},
  \bibinfo{author}{\bibfnamefont{A.}~\bibnamefont{Wacker}}, \bibnamefont{and}
  \bibinfo{author}{\bibfnamefont{E.}~\bibnamefont{Sch\"oll}},
  \bibinfo{journal}{cond-mat/0303025}  (\bibinfo{year}{2003}{\natexlab{b}}).

\bibitem[{\citenamefont{Blanter and B\"uttiker}(1999)}]{blanter1999}
\bibinfo{author}{\bibfnamefont{Y.~M.} \bibnamefont{Blanter}} \bibnamefont{and}
  \bibinfo{author}{\bibfnamefont{M.}~\bibnamefont{B\"uttiker}},
  \bibinfo{journal}{Phys. Rev. B} \textbf{\bibinfo{volume}{59}},
  \bibinfo{pages}{10217} (\bibinfo{year}{1999}).

\end{thebibliography}
\end{document}